\documentclass[preprint,12pt]{elsarticle}
\usepackage{lineno,hyperref}
\modulolinenumbers[1]
\usepackage{geometry}
\usepackage{graphicx}
\usepackage[usenames,dvipsnames]{color}
\usepackage{color}
\usepackage[english]{babel}
\usepackage{epstopdf}
\usepackage{latexsym}
\usepackage{hyperref}
\usepackage{url}
\usepackage{amstext}
\usepackage{amssymb}
\usepackage{amsmath}
\usepackage{pifont}
\usepackage{siunitx} 
\usepackage{url}
\hypersetup{
    colorlinks=true,
    linkcolor=blue,
    filecolor=magenta,
    urlcolor=blue,
}
\usepackage{times}
\usepackage{hyperref}
\usepackage{url}
\usepackage{amstext}
\usepackage{amssymb}
\usepackage{wrapfig}
\usepackage{float}
\usepackage{tikz}
\usepackage{caption}
\usepackage{subcaption}
\usepackage[section]{placeins}
\biboptions{sort&compress}

\begin{document}

\begin{frontmatter}

\title{Complex non-Markovian dynamics and the dual role of astrocytes in Alzheimer's disease development and propagation}

\author[inst1]{Swadesh Pal\corref{cor1}}
\ead{spal@wlu.ca}
\author[inst1,inst2]{Roderick Melnik}
\ead{rmelnik@wlu.ca}

\cortext[cor1]{Corresponding author}
\address[inst1]{MS2Discovery Interdisciplinary Research Institute, Wilfrid Laurier University, Waterloo, Canada}
\address[inst2]{BCAM - Basque Center for Applied Mathematics, E-48009, Bilbao, Spain}

\begin{abstract}
Alzheimer's disease (AD) is a common neurodegenerative disorder nowadays. Amyloid-beta (A$\beta$) and tau proteins are among the main contributors to the AD progression. In AD, A$\beta$ proteins clump together to form plaques and disrupt cell functions. On the other hand, the abnormal chemical change in the brain helps to build sticky tau tangles that block the neuron's transport system. Astrocytes generally maintain a healthy balance in the brain by clearing the A$\beta$ plaques (toxic A$\beta$). However, over-activated astrocytes release chemokines and cytokines in the presence of A$\beta$ and react to pro-inflammatory cytokines, further increasing the production of A$\beta$. In this paper, we construct a mathematical model that can capture astrocytes' dual behaviour. Furthermore, we reveal that the disease progression depends on the current time instance and the disease's earlier status, called the ``memory effect''. We consider a fractional order network mathematical model to capture the influence of such memory effect on AD progression. We have integrated brain connectome data into the model and studied the memory effect, the dual role of astrocytes, and the brain's neuronal damage. Based on the pathology, primary, secondary, and mixed tauopathies parameters are considered in the model. Due to the mixed tauopathy, different brain nodes or regions in the brain connectome accumulate different toxic concentrations of A$\beta$ and tau proteins. Finally, we explain how the memory effect can slow down the propagation of such toxic proteins in the brain, decreasing the rate of neuronal damage. 
\end{abstract}


\end{frontmatter}

\section{Introduction}{\label{intro}}

Alzheimer's disease (AD) is a neurological disorder that worsens with age and is incurable. It affects thinking, memory, and behaviour. These cognitive declines may be so severe that they interfere with daily tasks. Dr. Alois Alzheimer first observed this disease in 1906 and described it as ``a peculiar disease'' \cite{alzheimer1906}. He studied the brain of a lady who had passed away from an uncommon mental condition that included memory loss and linguistic issues. He found many abnormal clumps (amyloid-beta plaques) and fibre bundle tangles (tau tangles), which are now considered as one of the main contributors to AD progression \cite{alzheimer1906, moller1998, hippius2003}. They block communication between nerve cells, disrupt many processes, and cause memory loss, difficulty in speaking and other cognitive declines. 

Alzheimer's disease is not a natural part of the ageing process, although the chance of developing it grows with age \cite{harman2006,sengoku2020}. The majority of Alzheimer's patients are 65 years or older, although the disease can develop before that age, a condition known as early-onset illnesses AD \cite{bagyinszky2014}. Early in the course of the disease, people with Alzheimer's disease have little memory loss, but as they become older, they progressively lose their ability to maintain focus on a conversation or recall familiar faces. AD patients can survive up to twenty years following diagnosis, although the average is eight to ten years \cite{grossberg2003, weiner2005}. AD does not yet have a treatment, except a few drugs, such as aducanumab, which can assist in slowing cognitive loss in the early stages \cite{sevigny2016, selkoe2019, howard2020}. Many researchers have been working worldwide to understand the disease in a better way and prevent it from spreading.

Toxic amyloid-beta (A$\beta$) accumulation in the extracellular space is commonly regarded as one of the key initiators of the early start of AD \cite{murpy2010, gouras2015, liu2022role, rischel2022, roda2022}. This accumulation may happen due to its overproduction or the lower clearance rate \cite{sun2015}. A$\beta$ comprises 39-43 amino acids with different biophysical states, and soluble A$\beta_{40}$ and insoluble A$\beta_{42}$ are the two major isoforms observed in the brain. In a healthy brain, over $90\%$ concentration of A$\beta$ is detected in the form of A$\beta_{40}$, whereas less than $5\%$ can be found in the form of A$\beta_{42}$ \cite{burdick1992,gravina1995,kim2007}. Unfortunately, an AD-affected brain has no ability to maintain this state of balance and higher levels of A$\beta_{42}$ peptide lead plaques to develop, disrupting cell function. In addition, tau protein ($\tau$P) has a crucial function in AD \cite{medeiros2011,muralidar2020}. The normal $\tau$P creates a microtubule that helps transmit nutrients and other substances from one area of a nerve cell to another \cite{guo2017}. Abnormal chemical changes in the brain cause tau protein to separate from microtubules and attach to other $\tau$P \cite{kargbo2019,illenberger1998}. This causes the tau protein to form neurofibrillary tangles (misfolded and abnormally shaped) inside neurons and block the neuron's transport system. 

Researchers have focused on identifying toxic amyloid-beta and tau protein concentrations at the early onset of AD. At present, the accumulation of these proteins has not been completely quantified using blood tests and cannot be observed on CT or MRI images. The FDA-approved amyloid PET scan tracer can identify the existence of Alzheimer's disease, but it cannot adequately monitor disease progression; therefore, it is only used in clinical studies \cite{johnson2013,grundman2016,rabinovici2019}. In contrast, F-18 flortaucipir is the first FDA-approved tau PET scan tracer that aids with the progression of AD neurodegeneration \cite{devous2018,devous2021}. Along with these two proteins, many other factors influence AD progression. Substantial efforts have been made to identify the disease state based on different factors \cite{odagaki2017, parks2024}. 

A specific form of glial cell is found in the central nervous system (CNS), which serves as an immunodefense to the CNS. They control blood flow, transport mitochondria to neurons, and contribute to neuronal metabolism \cite{abbott2006,eroglu2010,hayakawa2016,kim2019}. Active astrocytes generally clean waste from the brain and protect neurons against illness \cite{tasdemir2014,koizumi2018,pal2022a,yuan2022}. But, in the AD-affected brain, they lose the ability to maintain a healthy balance and support AD progression \cite{garwood2017}. In the early stages of AD, a sufficient amount of toxic amyloid-beta mainly disrupts this healthy balance. In this circumstance, astrocytes cannot maintain the brain's ionic equilibrium, particularly intracellular Ca$^{2+}$ concentrations. As a result, NADPH oxidase (NOX) is activated, and neuronal death occurs due to oxidative stress \cite{abramov2005,kim2018}. Many other detrimental repercussions of astrocyte overactivation occur, including apolipoprotein E (ApoE) and excessive glutamate production. ApoE4 is a neurotoxic isoform of ApoE that contributes to toxic A$\beta$ deposition during the early stages of Alzheimer's disease \cite{bagyinszky2014,liu2017}. 

One of the main aspects of our current work is to analyze astrocytes' dual role before and after AD. We developed a mathematical model that incorporates the involvement of astrocytes in AD along with the A$\beta$ and $\tau$P interactions. In reaction kinetics, each of these proteins (A$\beta$ and $\tau$P) follows a heterodimer model for interactions between proteins, with a coupling parameter between them \cite{thompson2020}. We modify the exponential growth by logistic growth in the growth term for both the healthy proteins' equation \cite{meisl2021,pal2022}. We consider a logistic expansion in the astrocyte equation and presume they eliminate toxic amyloid-beta \cite{tacconi1998,thal2012}. Furthermore, these toxic proteins damage the neurons in the brain connectome. The amount of neuronal damage is studied here by coupling toxic A$\beta$ and $\tau$P, which also gives the disease status in the brain \cite{thompson2020}. 

A key challenge in modelling complex biological systems is extracting meaningful insights from available data. To address this, researchers have explored various modelling approaches that enhance interpretability and predictive power. Fractional calculus offers significant advantages over traditional integer-order models, as it naturally captures non-Markovian dynamics. Due to its global correlation, it can reflect the historical process of the systematic function and act as nonlocal interactions. Regarding the data fitting, it has been observed that the fractional model has one more degree of freedom over the traditional integer-order model \cite{chen2021}. Considering that a reaction-diffusion process could rely not only on the previous time instance's concentrations but also on each of the past stages of concentrations with specified weights, which is further discussed in this study \cite{stanislavsky2000, cressoni2012, cressoni2013, saeedian2017, troparevsky2019, mohammad2021}. 

Fractional-order derivative models widen classical calculus by expanding differentiation to non-integer orders, including memory effects and long-range interactions common in complex systems. They are specified using integral formulas like the Riemann-Liouville and Caputo derivatives, making them useful for simulating anomalous diffusion, viscoelastic materials, and biological processes \cite{caputo1967,troparevsky2019, vilk2024}. Fractional operators, such as the fractional Laplacian, use integrals over entire domains to capture non-local behaviour when extended to subsets of Euclidean space, often requiring specialized boundary conditions. This framework enables the formulation of fractional partial differential equations on restricted domains, which are commonly applied in fields like mathematical biology, image processing, and control theory, where classical integer-order models may fall short in representing complex spatial-temporal dynamics.

A time-fractional reaction-diffusion equation is often used to explore the memory effects in AD processes, as it can capture the influence of past states on the present dynamics. Investigating such memory effects in AD patients is crucial, as the disease progression and recovery rates can vary significantly between individuals. For instance, the recovery rate of a patient who has been affected by the disease for twenty years may differ from that of a patient who has been diagnosed for only ten years, highlighting the need for models that account for long-term memory and individual variability in disease dynamics. There is a growing number of works on fractional models, which are applied to mathematical biology and other areas, such as fractional reaction-diffusion models in pattern formation and the dynamics of chemical kinetics in a heterogeneous setting \cite{burrage2024}, a typical measure of reaction time in thermally activated barrier-crossing processes \cite{zhou2024}, and long-range movement of certain organisms in the presence of a chemoattractant \cite{estrada2018}, gene expression \cite{vilk2024}, cell motions \cite{mitterwallner2020}, long-range memory \cite{de2022}, etc. With several additional practical uses, the Caputo derivative is among the best fractional operators for use in this kind of modelling \cite{saeedian2017, carvalho2018, ghosh2021}. In this work, we first construct a time-fractional partial differential equation (PDE) model to describe the AD progression. Then, we develop the network model that aligns with the PDE model to integrate the data on brain connections and examine the damage dynamics associated with the influence on memory. Different tauopathies have been studied for the network model to incorporate different scenarios depending on the toxic A$\beta$ and toxic $\tau$P. Furthermore, we have compared the disease progression in the absence and presence of memories on nodes and brain connectome regions.

The rest of this paper is organized as follows. In Sect. \ref{sec2}, we formulate the temporal models for AD progression for both cases: absence and presence of memories. The equilibria and their stabilities for the temporal model are discussed in Sect. \ref{sec3}. In Sect. \ref{sec4}, the temporal model is enlarged inside a subset of Euclidean space and then incorporated into the network to include brain connectome data. Comprehensive numerical simulation outcomes are displayed in Sect. \ref{sec5} to analyze the dual role of astrocytes and the memory effect in AD progression. Finally, outcomes and future prospects are discussed in Sect. \ref{sec6}.


\section{Results and Discussions}{\label{sec5}}

\begin{figure}[ht!]
\centering
\includegraphics[width=\textwidth]{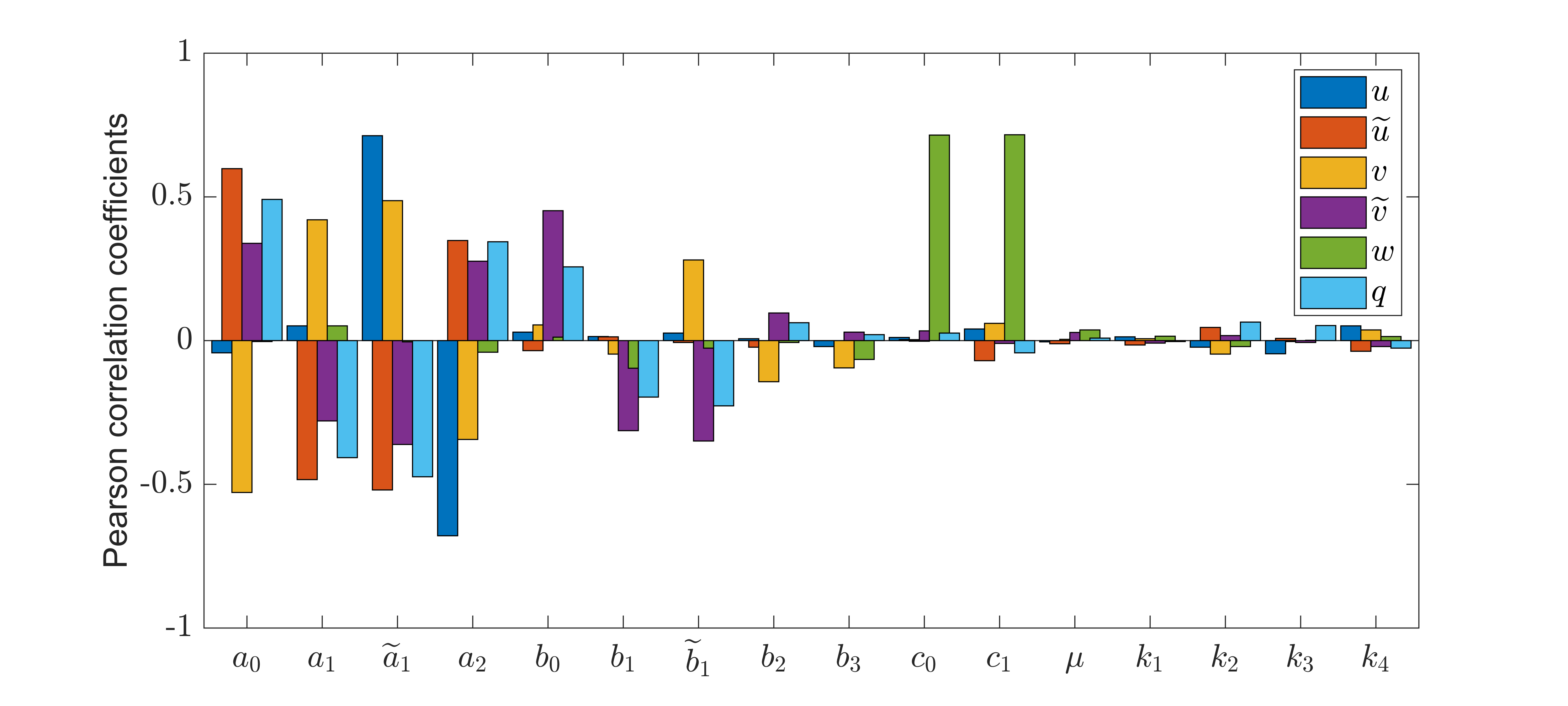}
\caption{(Color online) The sensitivity analysis of different parameters on the non-fractional model of (\ref{MHDMPL}) and (\ref{TDE}).}
\label{fig:figs1}
\end{figure}

This section presents the numerical results for the non-fractional and fractional models applied to the brain connectome. Before conducting numerical simulations, we introduce the synthetic parameter values used in the models, listed in Table \ref{table:tab2}, as estimated by Thompson et al. \cite{thompson2020}. Additionally, we have performed a sensitivity analysis, which is essential for assessing how variations in parameter values affect the model’s behaviour. This analysis helps identify key parameters that drive system dynamics, emphasizing those with a significant impact on the outputs, which may require precise estimation or experimental validation. Figure \ref{fig:figs1} illustrates the sensitivity analysis results for the non-fractional model of (\ref{MHDMPL}) and (\ref{TDE}). Pearson correlation coefficients were calculated for each parameter by generating one thousand uniform random samples with a 10\% deviation from the values listed in Table \ref{table:tab2} and evaluating the model's solution at $t=200$. In this case, the initial condition is chosen as $u_{0} = 0.75$, $\widetilde{u}_{0} = 0.0075$, $v_{0} = 0.5$, $\widetilde{v}_{0} = 0.005$, $w_{0} = 0.5$, and $q_{0} = 0$. The bar plot reveals that parameter $a_{0}$ exhibits a strong positive correlation with $\widetilde{u}$, $\widetilde{v}$, and $q$, while it shows a strong negative correlation with $v$ and a weaker correlation with both $u$ and $w$ variables. Sensitivity patterns for the other parameters are also illustrated in the figure. Overall, the parameters on the left side of the x-axis label in Fig. \ref{fig:figs1} exhibit stronger correlations with $q$. Therefore, they may serve as potential targets for AD-modifying therapies.

\begin{table}
\caption{Synthetic parameter values \cite{thompson2020}.}
\label{table:tab2}
\begin{center}
\begin{tabular}{|c|c|c|c|c|c|}
\hline
Parameter & Value & Parameter & Value & Parameter & Value\\
\hline
$a_{0}$ & 1.035 & $a_{1}$ & 1.38 & $a_{2}$ & 1.38 \\
\hline
$\widetilde{a}_{1}$ & 0.828 & $b_{0}$ & 0.69 & $b_{1}$ & 1.38 \\
\hline
$b_{2}$ & 1.035 & $\widetilde{b}_{1}$ & 0.552 & $b_{3}$ & 4.14 \\
\hline
$c_{0}$ & 1.0& ${c}_{1}$ & 0.1 & $\mu$ & 0.1 \\
\hline
$\rho_{1}$ & 1.38 & $\rho_{2}$ & 0.138 & $\rho_{3}$ & 1.38 \\
\hline
$\rho_{4}$ & 0.014 & $k_{1}$ & 0.0001 & $k_{2}$ & 0.01 \\
\hline
 $k_{3}$ & 0.1 & $k_{4}$ & 0.001 & & \\
\hline
\end{tabular}
\end{center}
\end{table}

We use the brain connectome data, which is available freely at BrainGraph.org – the network of the brain. These data give information on a network with nodes and edges in various brain areas, allowing us to investigate the brain's spatio-temporal behaviour. In this brain graph data, each node corresponds to a tiny area ($1-1.5 cm^{2}$) of the gray matter, called the region of interest (ROI). An edge may be connected to two nodes if a diffusion-MRI-based procedure discovers fibers of axons going between those two nodes in the brain's white matter \cite{kerepesi2016,szalkai2019}. We have integrated the brain connectome data into our computational environment (Matlab) and extracted the corresponding Laplacian for the real data. The network data consists of 1,015 nodes and 16,280 edges. The number of fibers in the integrated data varies between 1 and 4,966.5, with an average of 39.33 fibers per edge. Most of these fibers are located between the superior parietal and precuneus regions \cite{shaheen2023a}. Additionally, the average fiber length ($l_{ij}$) ranges from 10.270 mm to 83.003 mm, with an average of 30.089 mm. We use the Laplacian to derive the numerical solution for the network model to handle various scenarios. The fourth-order Runge-Kutta method is applied to integrate the resulting system with a time step of $dt = 0.01$, and the results remain consistent for smaller time steps. Furthermore, the predictor-corrector method is employed to solve the fractional model \cite{diethelm2002}. We have computed the numerical results based on our code implemented in C-language. We have used Sharcnet (www.sharcnet.ca) supercomputers to run multiple jobs simultaneously, which helped us efficiently analyse different aspects of the model.

\begin{figure}[ht!]
\centering
\includegraphics[width=8cm]{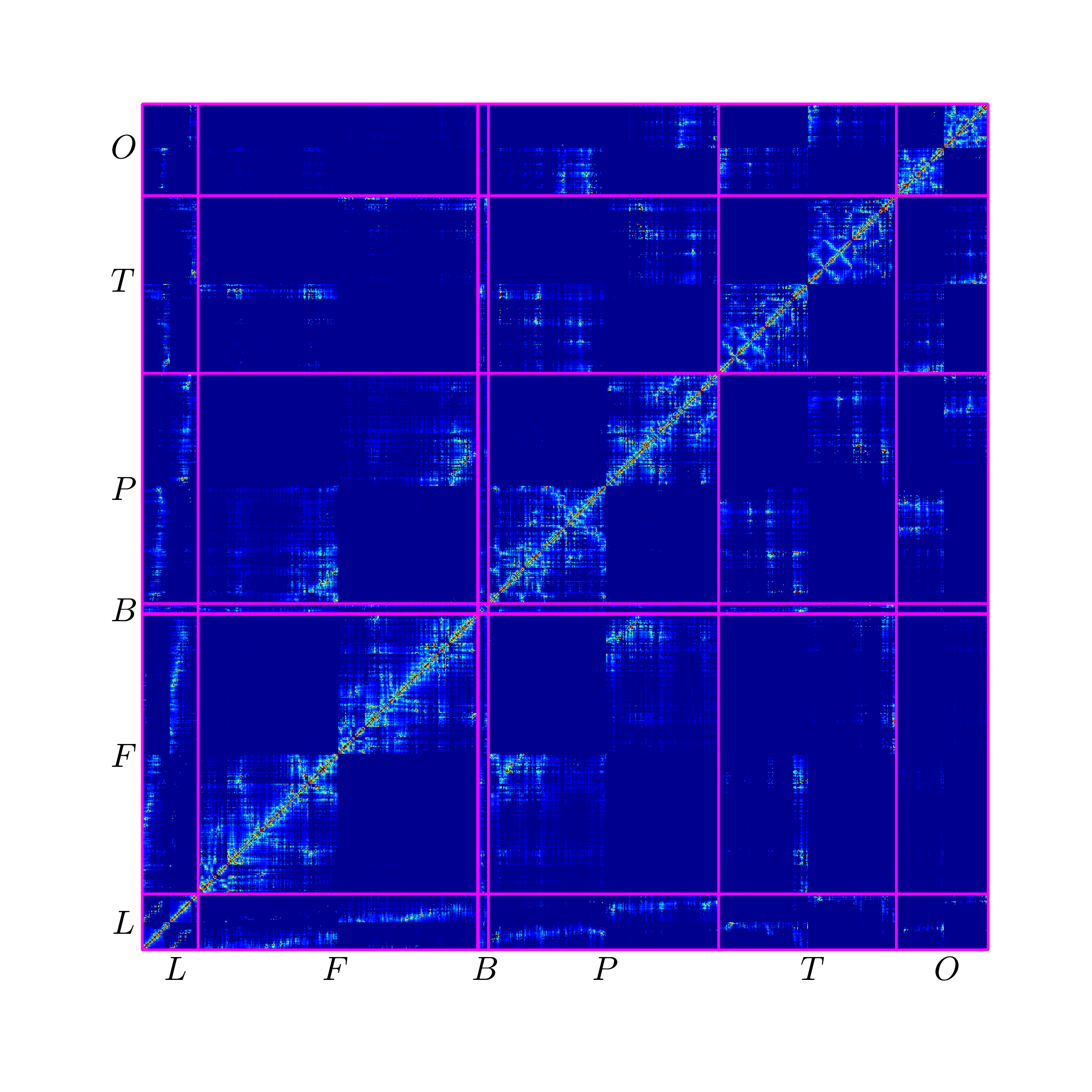}
\caption{(Color online) Weighted adjacency matrix for the brain connectome data: limbic ($L$), frontal ($F$), basal ganglia ($B$), parietal ($P$), temporal ($T$) and occipital ($O$). In our integrated brain connectome data, each region contains one or more brain IDs, and they are listed in Table \ref{table:tab1}.}
\label{fig:fig1}
\end{figure}

Figure \ref{fig:fig1} depicts the weighted adjacency matrix for the considered network data.
\begin{table*}
\caption{Brain IDs associated with brain regions \cite{fornari2020}.}
\label{table:tab1}
\begin{tabular}{|l|p{0.8\textwidth}|}
\hline
Brain region & Brain ID \\
\hline
 Limbic & Rostralanteriorcingulate, Posteriorcingulate, Caudalanteriorcingulate, Parahippocampal, Isthmuscingulate, Entorhinal \\
 \hline
 Frontal & Frontalpole, Lateralorbitofrontal, Parsorbitalis, Medialorbitofrontal, Precentral, Parstriangularis, Parsopercularis,  Rostralmiddlefrontal, Caudalmiddlefrontal, Superiorfrontal\\
 \hline
Parietal & Postcentral, Supramarginal, Superiorparietal, Inferiorparietal, Precuneus, Paracentral \\
\hline
Basal Ganglia & Left-Thalamus-Proper, Left-Putamen, Left-Caudate, Left-Accumbens-area, Left-Pallidum, Left-Amygdala, Right-Thalamus-Proper, Right-Putamen, Right-Caudate, Right-Accumbens-area, Right-Pallidum, Right-Amygdala \\
\hline
Brain Stem & Brain-Stem \\
\hline
Occipital & Cuneus, Pericalcarine, Lateraloccipital, Lingual \\
\hline
Temporal & Middletemporal, Left-Hippocampus, Right-Hippocampus, Temporalpole, Inferiortemporal, Bankssts, Superiortemporal, Transversetemporal, Insula, Fusiform \\
\hline
\end{tabular}
\end{table*}
In the plot, blue to red colours represent the minimum to maximum strength of the connection between the nodes. The integrated brain connectome data contains one or more brain IDs in each region, listed in Table \ref{table:tab1}. The brain stem region is also in the diagram between the frontal and basal ganglia, but it is not visible because it contains only one node. This figure depicts the relationship between the brain connectome's nodes and regions. This matrix determines the spatiotemporal behaviour of the brain connectome network model. According to the parameter values, both models (non-fractional and fractional models) share the feasible homogeneous steady-states. In the homogeneous steady-state, the concentration of toxic amyloid-beta could governed by the concentration of toxic tau protein. In this case, it is called secondary tauopathy; otherwise, it is a primary tauopathy. We will discuss both cases in the coming subsection.

Before moving to the numerical simulations, we first mention the initial conditions for each variable in the network model. In the brain connectome, the initial seeding sites for the toxic amyloid-beta are the temporobasal and frontomedial regions, and the toxic tau proteins are the transentorhinal and locus coeruleus associated regions \cite{thompson2020,pal2022}. For the seeding locations, we add tiny toxic concentrations of $0.25\%$ and $0.38\%$ in toxic tau protein ($\widetilde{v}$) and toxic amyloid-beta ($\widetilde{u}$), respectively. Due to these small perturbations, the toxic concentrations propagate all over the brain connectome and spread AD. On the other hand, we consider healthy concentrations for both amyloid-beta ($u$) and tau proteins ($v$) and a small concentration for the astrocytes ($w$). Some other perturbations of these initial concentrations can change the initial propagation profiles of the concentrations, but the final results (long-term behaviours) are the same. These concentrations are uniform on each node in the brain network. We set the initial condition $q = 0$ for the damage equation to each node. As toxic loads propagate over the brain connectome, they damage the neurons in the brain.

\subsection{Primary and secondary tauopathies}

\begin{figure}[ht!]
\begin{center}
\begin{subfigure}[p]{0.44\textwidth}
        \centering
       \includegraphics[width=\textwidth]{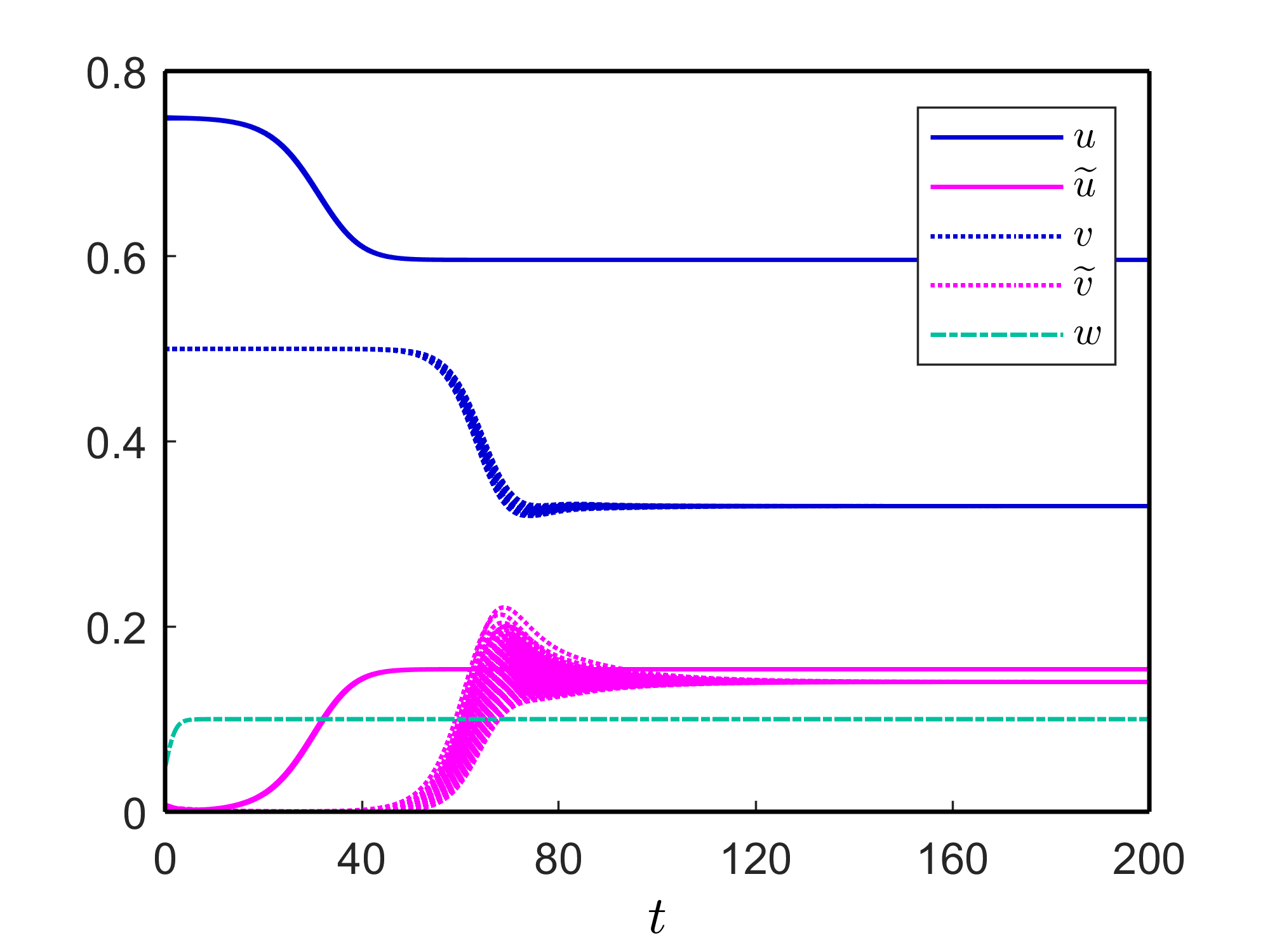}
        \caption{ }\label{fig:fig2a}
\end{subfigure}%
\begin{subfigure}[p]{0.44\textwidth}
        \centering
       \includegraphics[width=\textwidth]{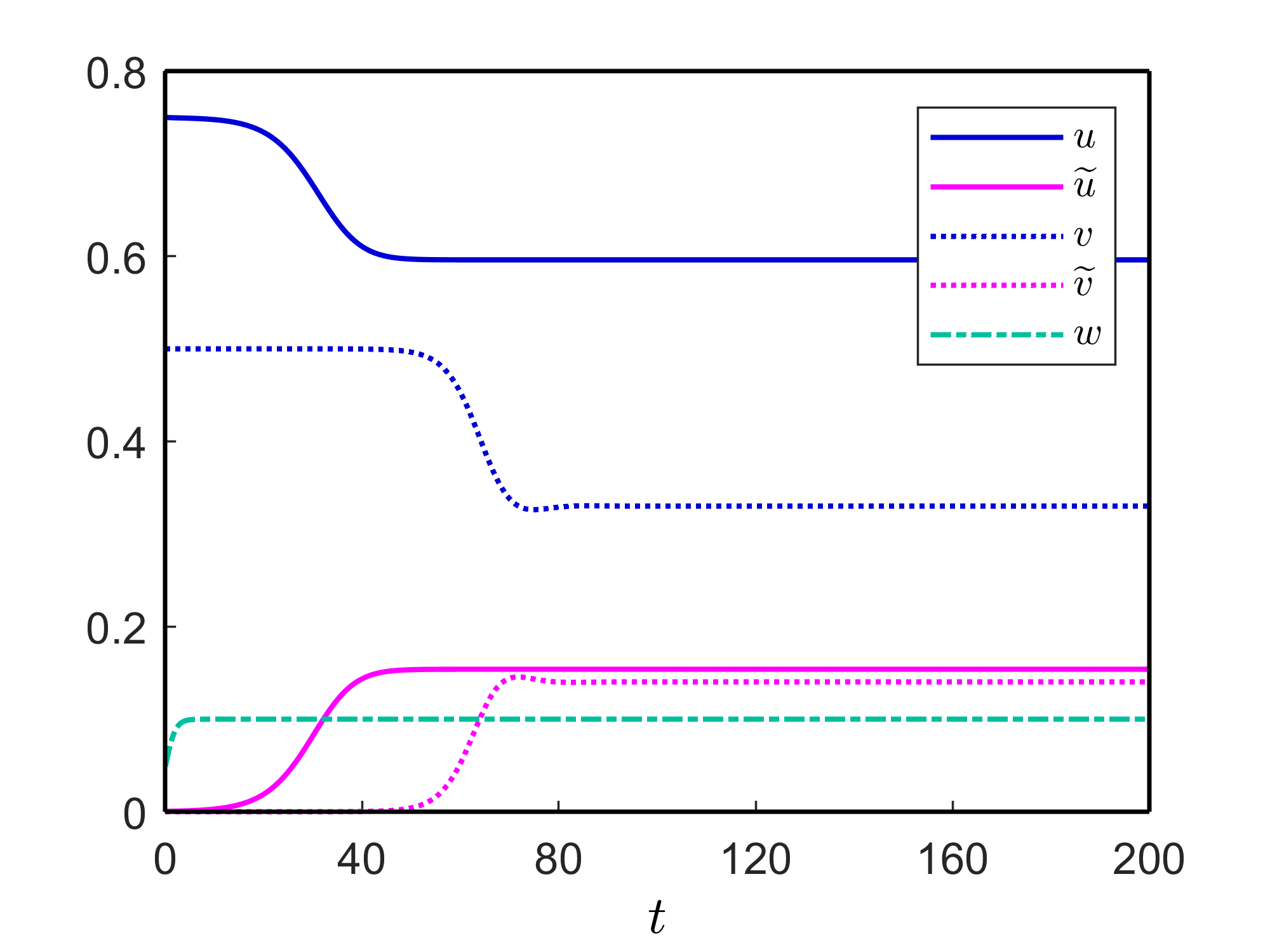}
        \caption{ }\label{fig:fig2b}
\end{subfigure}
\caption{ (Color online) Solutions for the non-fractional model of (\ref{NSMHDMPL}) in the brain connectome: (\subref{fig:fig2a}) solutions in each region and (\subref{fig:fig2b}) the spatial average solution. The fixed parameter values are given in Table \ref{table:tab2}.}\label{fig:fig2}
\end{center}
\end{figure}

As in our previous study, we have shown that the evolution profiles of both toxic loads remain the same for primary and secondary tauopathies in the absence of memory ($\alpha = 1$) and astrocytes \cite{pal2022}. We also observe that the profiles remain consistent for both tauopathies in the presence of astrocytes. Therefore, without any loss of generality, we present the results using the parameter values associated with the secondary tauopathy. Table \ref{table:tab2} provides a synthetic parameter set specific to the secondary tauopathy. For this parameter set, the non-trivial equilibrium point $E_{*} = (0.596,0.154,0.33,0.14,0.1)$ is locally asymptotically stable, and a numerical solution of the system is shown in Fig. \ref{fig:fig2}. Additionally, we have explored a more general scenario (mixed tauopathy), where non-uniform parameter values are assigned to different nodes in the brain connectome.

\subsubsection{Dual role of astrocytes}

\begin{figure}[ht!]
\begin{center}
        \begin{subfigure}[p]{0.44\textwidth}
                \centering
               \includegraphics[width=\textwidth]{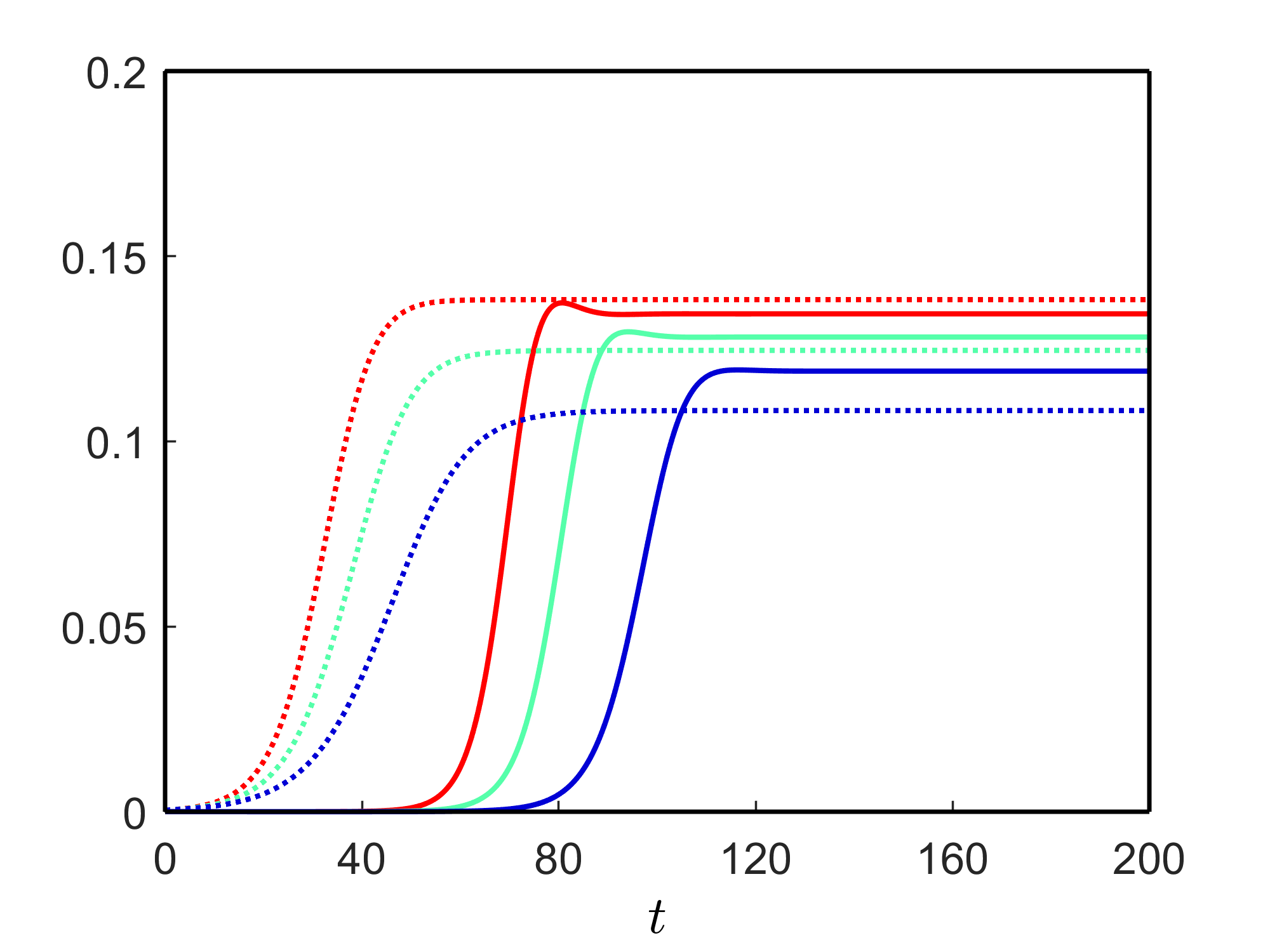}
                \caption{ $c_{1} = 0.3$}\label{fig:fig3a}
        \end{subfigure}%
        \begin{subfigure}[p]{0.44\textwidth}
                \centering
               \includegraphics[width=\textwidth]{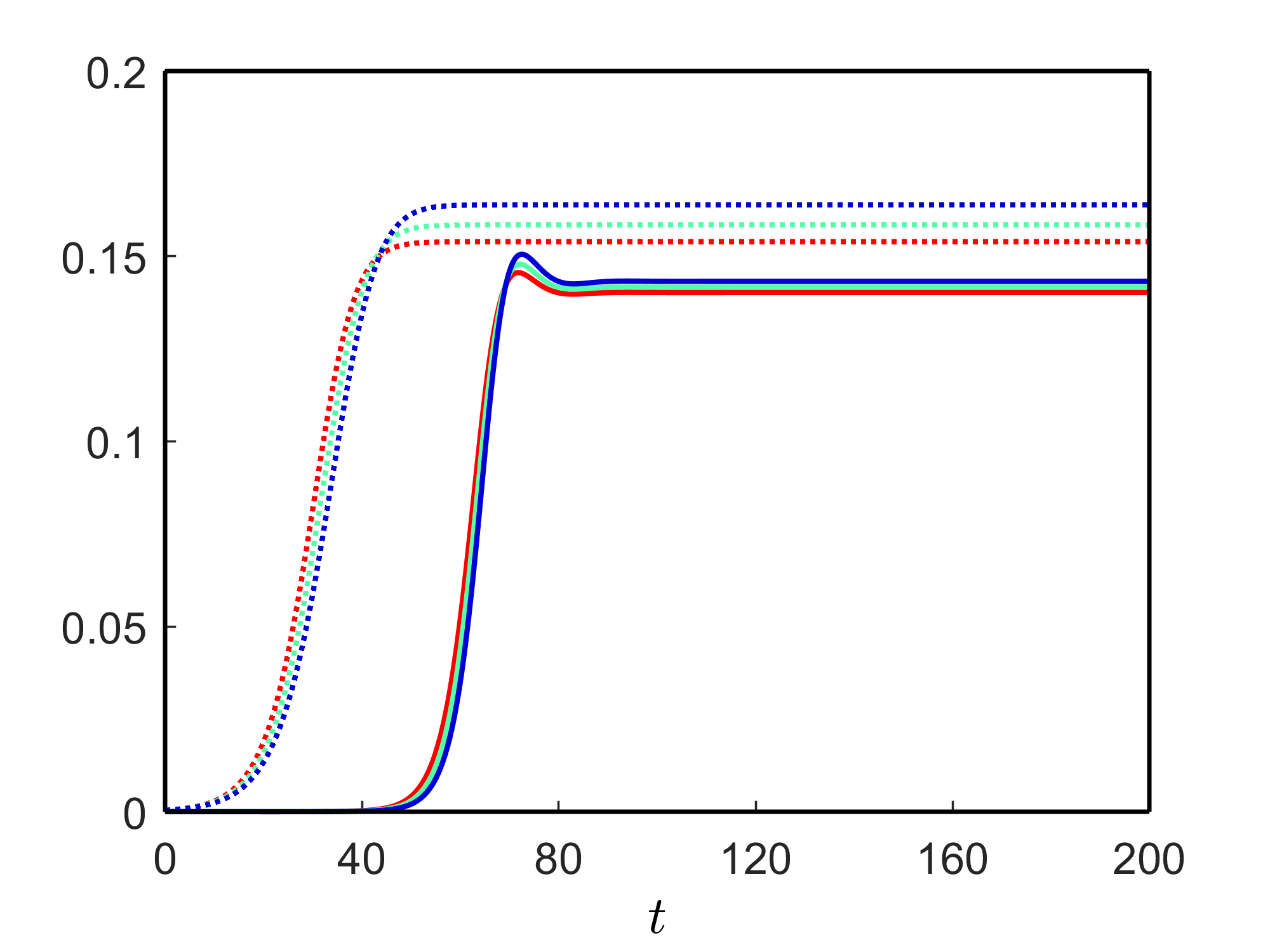}
                \caption{ $c_{1} = 0.1$ }\label{fig:fig3b}
        \end{subfigure}
        \caption{ (Color online) Spatial average solutions of toxic A$\beta$ (dotted) and toxic $\tau$P (solid) for the non-fractional model (\ref{NSMHDMPL}) for different values of $c_{1}$ and $\mu$: (red) $\mu = 0.1$, (green) $\mu = 0.2$ and (blue) $\mu = 0.3$. }\label{fig:fig3}
\end{center}
\end{figure}

As previously stated, astrocytes have a dual function in AD transmission. They work to remove harmful amyloid-beta and maintain a healthy equilibrium in the brain. However, due to the accumulation of toxic amyloid-beta, astrocytes become overactive and contribute to disease progression rather than brain protection. Therefore, two scenarios can occur depending on the concentrations of astrocytes present in the brain cells: they can manage the toxic amyloid-beta, but sometimes they cannot. Here, we capture both cases through our considered network mathematical model in the absence of memory. In our model, the parameter $c_{1}$ represents the brain cells' maximum concentration (carrying capacity) of astrocytes. We consider two different carrying capacities for the astrocytes in the network model, and the average toxic density propagations over time are shown in Fig. \ref{fig:fig3}. For $c_{1} =0.3$, increasing the clearance rate $\mu$ reduces the toxic burden on the brain connectome [see Fig. \ref{fig:fig3}(\subref{fig:fig3a})]. This shows that astrocytes can control the brain's toxic loads. On the other hand, $c_{1} =0.1$, they fail to manage the proper equilibrium in the brain connectome and encourage a rise in toxic loads [see Fig. \ref{fig:fig3}(\subref{fig:fig3b})]. The non-trivial equilibrium point $E_{*}$ is locally asymptotically stable in both cases.

\subsubsection{Memory effect}

\begin{figure}[ht!]
\centering
\includegraphics[width=8.3cm]{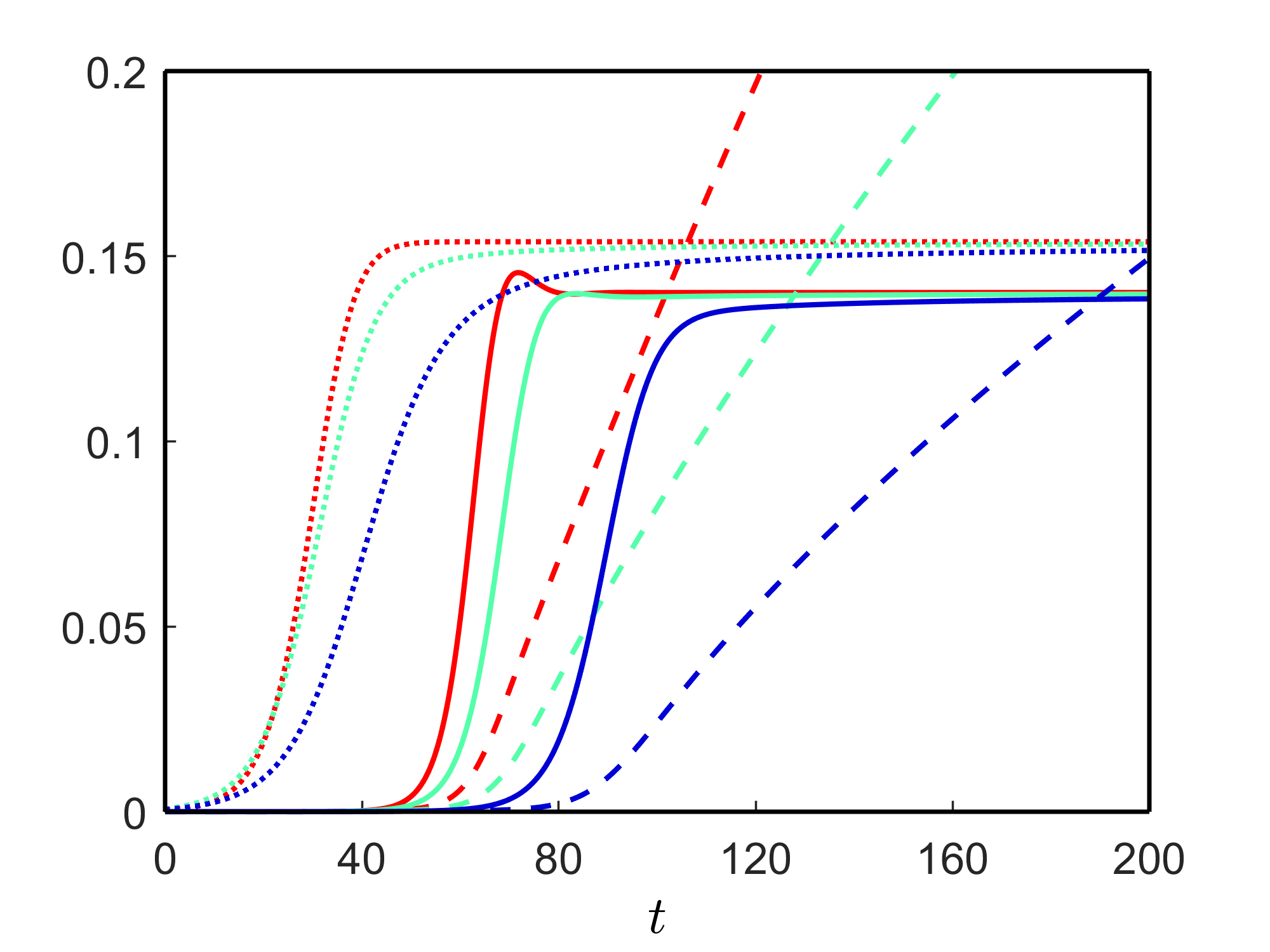}
\caption{(Color online) Spatial average solutions of toxic amyloid-beta (dotted), toxic tau protein (solid) and damage (dashed) for the fractional model of (\ref{NSMHDMPL}) with (\ref{NTDE}) for different values of $\alpha$ with the other fixed parameter values of Table \ref{table:tab2} over the brain connectome. In the plot, $\alpha = 1$, $\alpha = 0.9$, and $\alpha = 0.8$ are represented by the red, green and blue curves, respectively.}
\label{fig:fig4}
\end{figure}

Once memory effects become significant, the Markovian framework does not adequately describe the underlying complex dynamic processes behind the progression of neurodegenerative diseases. Here, we analyze the memory effect of AD progression in the brain connectome. As mentioned earlier, the model (\ref{NSMHDMPL}) has a memory for $0<\alpha < 1$ and memoryless for $\alpha \rightarrow 1$. As discussed in Sect \ref{sec2}, the underlying processes are non-Markovian. Figure \ref{fig:fig4} depicts both the toxic propagation over the brain connectome for no-memory and with memory. In the figure, we plot the spatial average of toxic amyloid-beta and toxic tau protein. In both cases ($\alpha = 0.9$ and $\alpha = 0.8$), the non-trivial equilibrium point $E_{*}$ satisfies the conditions for locally asymptotically stable, as mentioned in Sect \ref{sec3}.  Here, the evolution time of the toxic loads for the fractional model is higher compared to the non-fractional model. Furthermore, with an increase in the memory effect (decreasing the value of $\alpha$), the evolution time of toxic loads also increases [see Fig. \ref{fig:fig4}].

\subsubsection{Neuronal damage}

Following the model (\ref{NTDE}), the neuronal damage depends on the toxic concentrations present in the brain connectome; hence, the total brain damage depends on the evolutional time (the time required to converge to the stable steady state) of toxic loads. We have mentioned the parameter values directly associated with the neuronal damage in Table \ref{table:tab2}. These parameter values give us the influence of toxic tau proteins on neural damage and the presence of toxic amyloid-beta \cite{small2008,lloret2011,cho2016,jack2018}. We plot the spatial average of the damage in Fig. \ref{fig:fig4}, and it validates the dependency. It has been observed that the damage converges to its equilibrium point $q_{*} = 1$ for both fractional and non-fractional models, but the case of the fractional model takes a longer time than the non-fractional model. Overall, memory has a pronounced effect on AD progression.

\subsection{Mixed tauopathy}

Here, we focus on disease progression for non-uniform parameters over the brain connectome. This is more realistic than the uniform parameters as the presence of heterogeneous density of the ingredients in the brain (e.g., proteins, chemical ions, etc.). We consider the parameter values of $b_{2}$ and $b_{3}$ from Table \ref{table:tab2} in all the brain identities (IDs) except some regions mentioned in \cite{thompson2020,pal2022}. The methodology of getting these values is mentioned in \cite{thompson2020}. In the network model, a combination of primary and secondary tauopathies, known as mixed tauopathy, arises because of the non-uniform parameter values in the brain connectome. This causes different stable coexisting steady-states in the network model, and we divide these into two parts: region ID and region-wise disease progression.

\begin{figure}[ht!]
\begin{center}
        \begin{subfigure}[p]{0.44\textwidth}
                \centering
               \includegraphics[width=\textwidth]{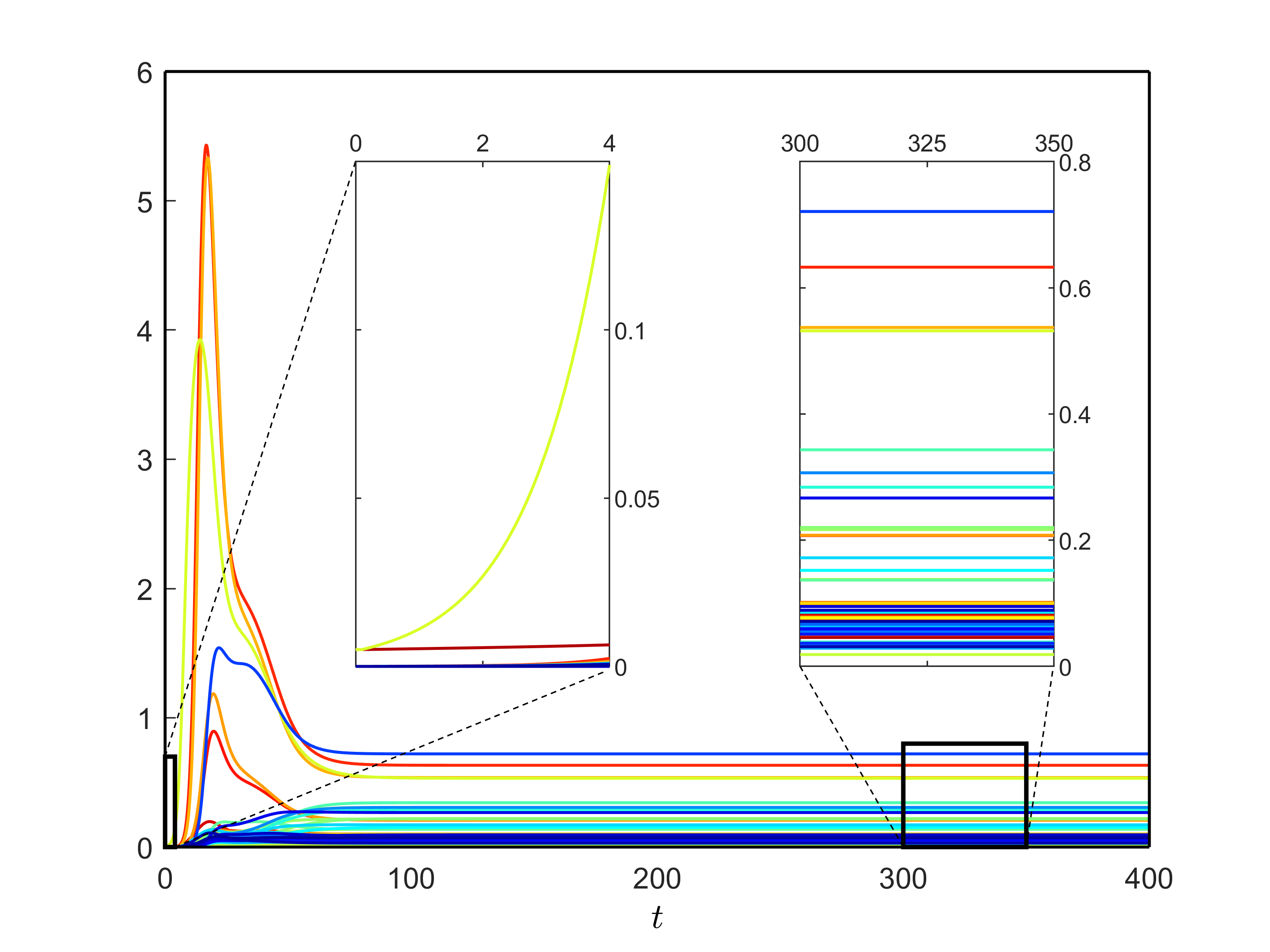}
                \caption{  }\label{fig:fig5a}
        \end{subfigure}%
        \begin{subfigure}[p]{0.44\textwidth}
                \centering
               \includegraphics[width=\textwidth]{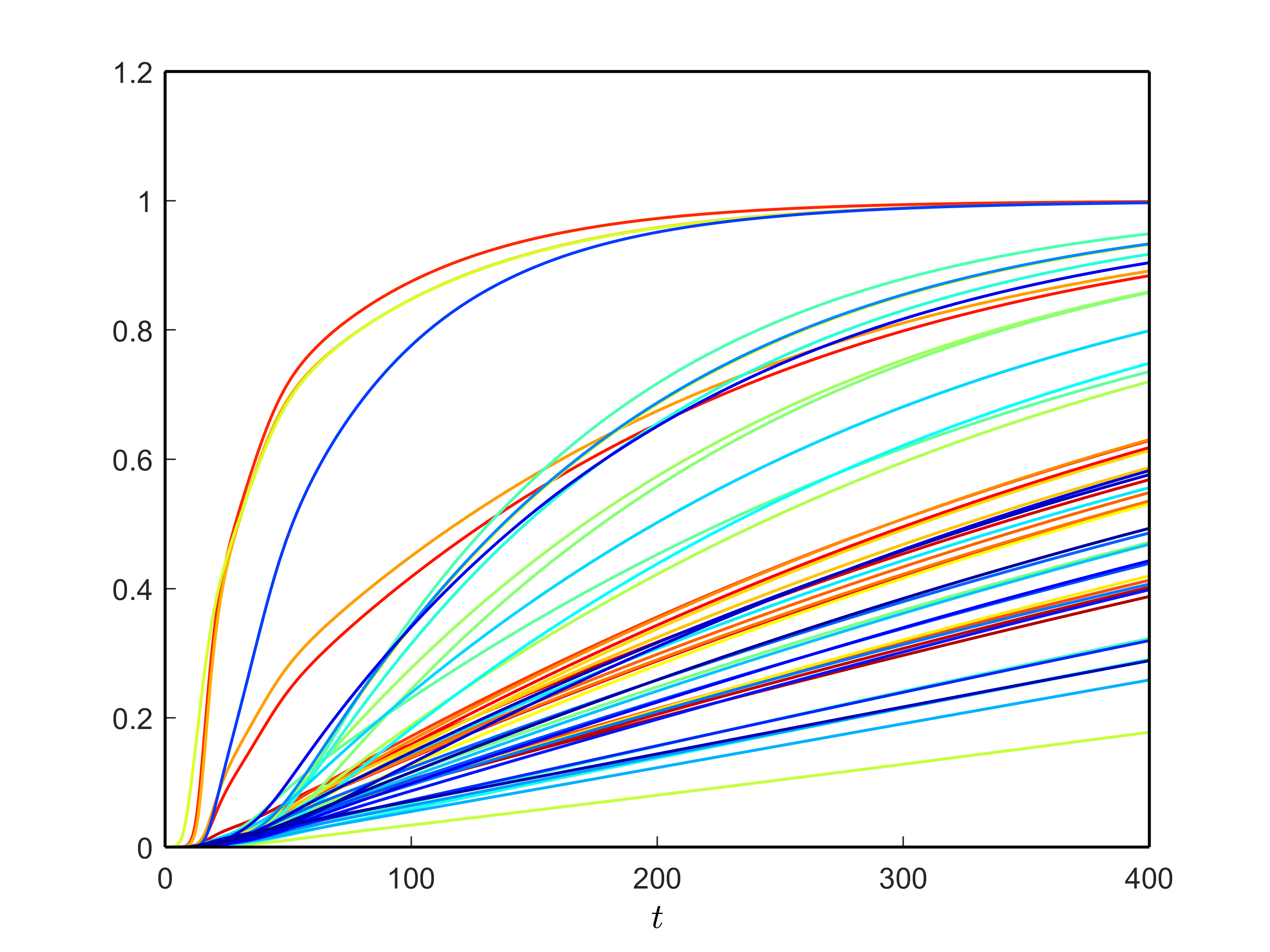}
                \caption{ }\label{fig:fig5b}
        \end{subfigure}
        \caption{(Color online) Brain ID-wise average toxic tau protein propagation (\subref{fig:fig5a}) and the corresponding neural damage (\subref{fig:fig5b}) for the non-fractional model. The $y$-axis represents the brain-ID-wise average toxic tau protein concentrations. }\label{fig:fig5}
\end{center}
\end{figure}

\subsubsection{Region ID-wise AD progression}

The integrated brain connectome data contains forty-nine brain IDs, each with one or more nodes. We calculate the average concentration of the toxic amyloid-beta for each brain ID by the formula \cite{fornari2020}:
\begin{equation}{\label{ACF}}
    M_{\widetilde{u}}^{d} = \frac{1}{n_{d}}\sum_{k\in \mathcal{R}_{d}}\widetilde{u}_{k},
\end{equation}
where $\mathcal{R}_{d}$ is the set of all nodes in that brain ID, and $n_{d}$ denotes the total number of elements in $\mathcal{R}_{d}$. We use the same formula for the toxic tau proteins and damage dynamics. For the non-fractional model, we observe uniform average concentrations of toxic amyloid-beta (not shown here) and non-uniform average concentrations of toxic tau proteins along the brain IDs [see Fig. \ref{fig:fig5} (\subref{fig:fig5a})]. This happens due to the direct involvement of the non-uniform parameters $b_{2}$ and $b_{3}$ in the healthy and toxic tau proteins equations. Furthermore, the damage propagation profiles for each brain ID are different [see Fig. \ref{fig:fig5} (\subref{fig:fig5b})]. According to the integrated brain connectome data, the maximum concentrations of toxic amyloid-beta accumulate in the region ID precuneus, followed by the region IDs left-putamen, right-putamen, entorhinal, etc. The damage dynamics show that these region IDs are affected the most at the initial stage of AD progression.  

\subsubsection{Region-wise AD progression}

\begin{figure}[ht!]
\begin{center}
\begin{subfigure}[p]{0.3\textwidth}
        \centering
       \includegraphics[width=\textwidth]{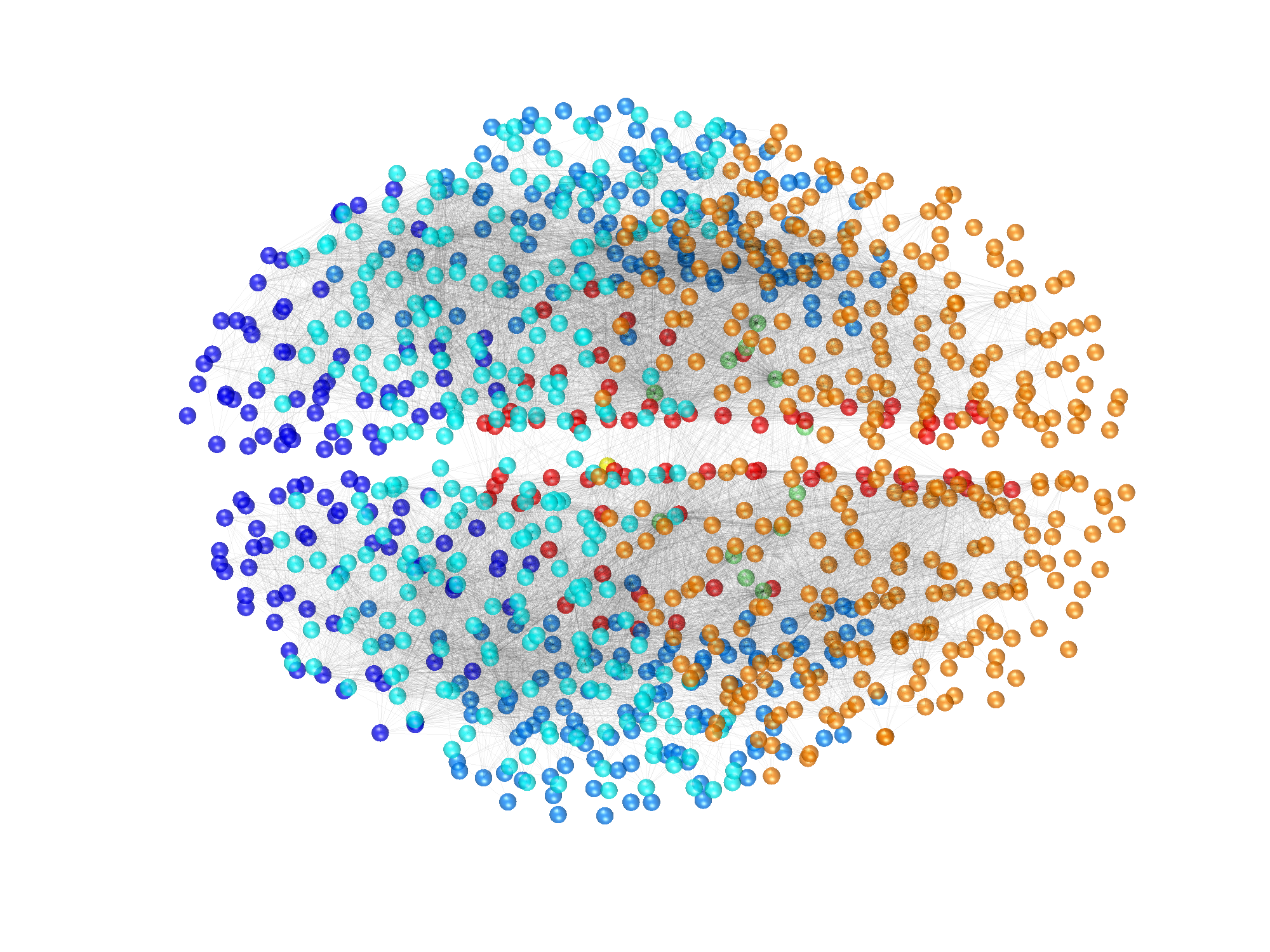}
        \caption{  }\label{fig:fig6a}
\end{subfigure}%
\begin{subfigure}[p]{0.3\textwidth}
        \centering
       \includegraphics[width=\textwidth]{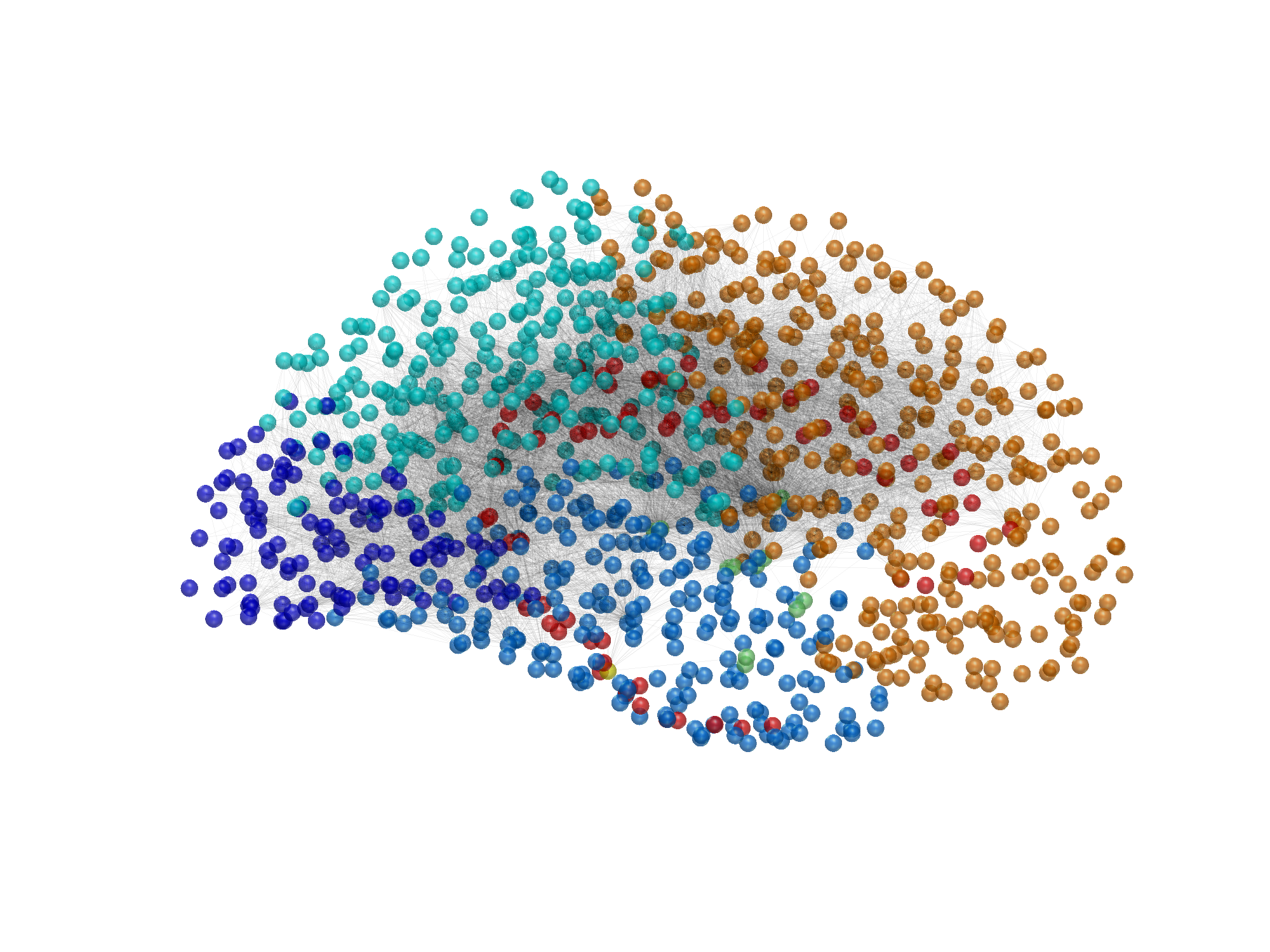}
        \caption{ }\label{fig:fig6b}
\end{subfigure}
\caption{(Color online) Three-dimensional views of the positions of the nodes for the integrated brain connectome data: (\subref{fig:fig6a}) axial view and (\subref{fig:fig6b}) sagittal view. Different colours are used to indicate different brain regions. }\label{fig:fig6}
\end{center}
\end{figure}

\begin{figure}[ht!]
\begin{center}
\begin{subfigure}[p]{0.44\textwidth}
        \centering
       \includegraphics[width=\textwidth]{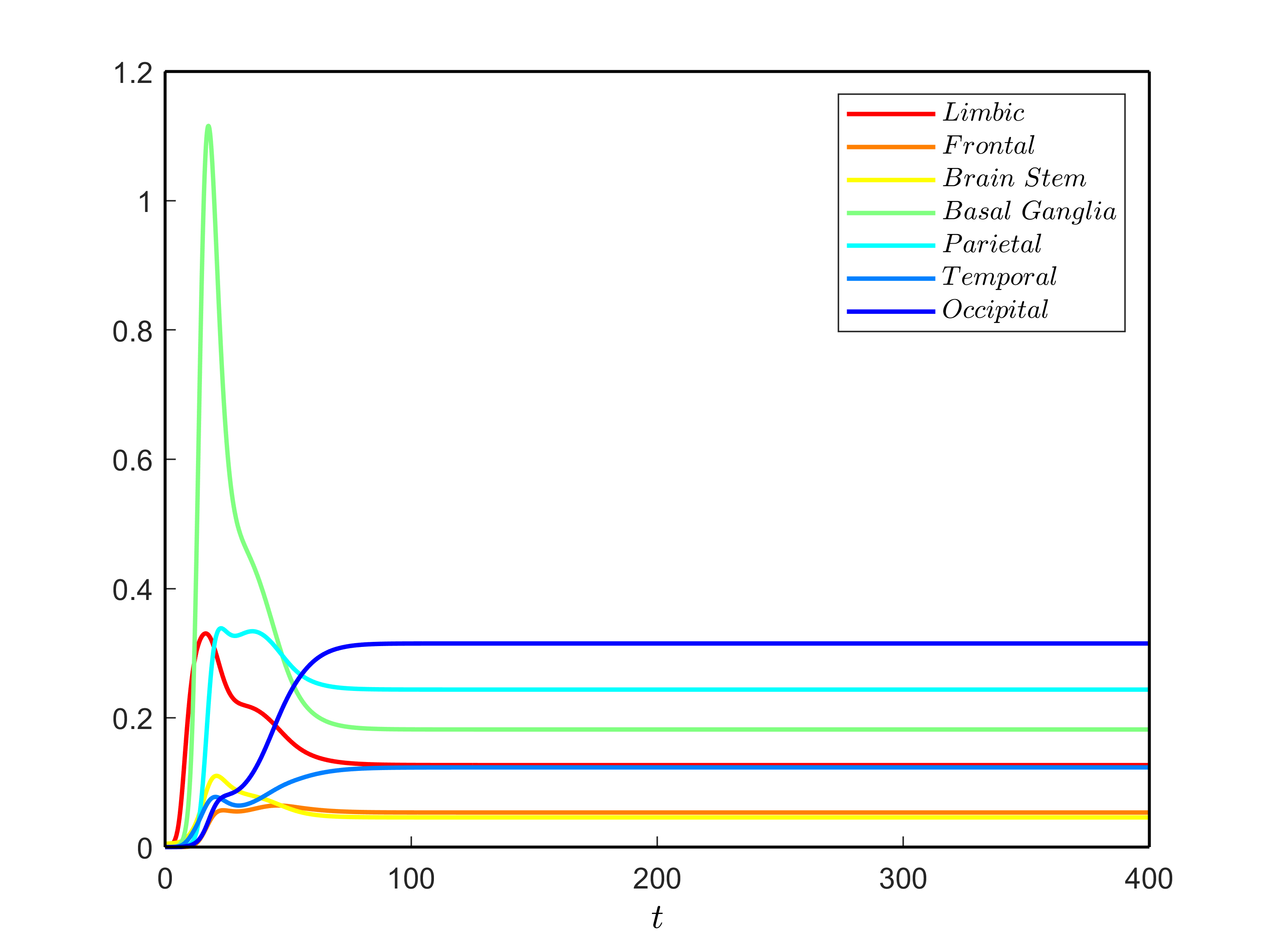}
        \caption{ }\label{fig:fig7a}
\end{subfigure}%
\begin{subfigure}[p]{0.44\textwidth}
        \centering
       \includegraphics[width=\textwidth]{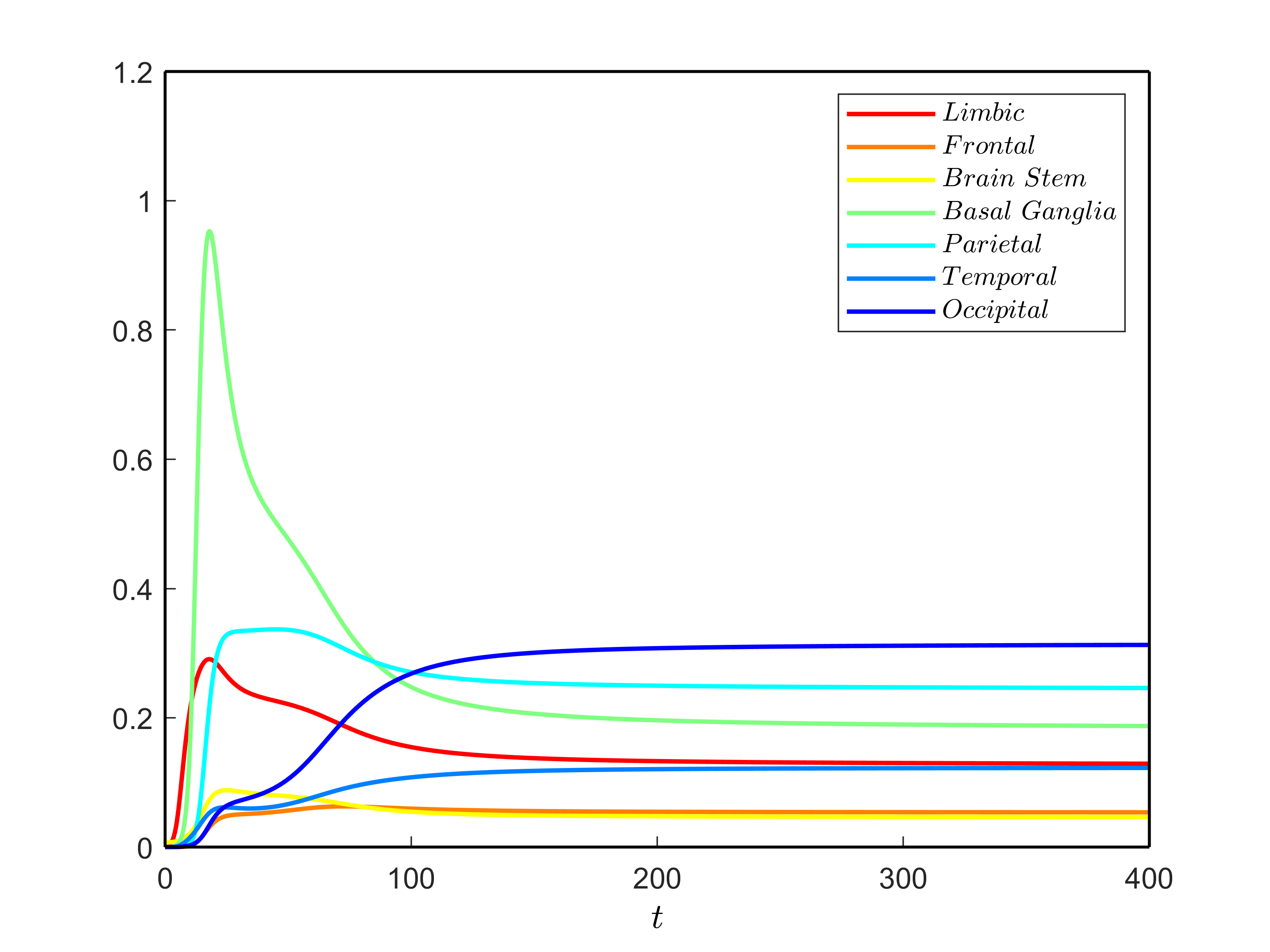}
        \caption{ }\label{fig:fig7b}
\end{subfigure}
\caption{ (Color online) Brain region-wise average toxic tau protein ($\widetilde{v}$) propagation for the non-fractional (\subref{fig:fig7a}) and fractional with $\alpha = 0.8$ (\subref{fig:fig7b}) models. The $y$-axis represents the average toxic tau protein concentrations. }\label{fig:fig7}
\end{center}
\end{figure}

We focus on the evolution of the toxic load distributions and their damage profile for seven brain regions (brain stem, frontal, temporal, limbic, basal ganglia, parietal, and occipital), and each region containing one or more brain IDs. The integrated brain connectome data contains Cartesian coordinates for all the nodes in three-dimensional space and their brain IDs. We plot them according to their regions (mentioned in Table \ref{table:tab1}) in Fig. \ref{fig:fig6}, and in the plot, different colours of the nodes belong to different regions. We have used the same colour codes in Figs. \ref{fig:fig6} and \ref{fig:fig7} for the brain regions.
    
We apply the formula (\ref{ACF}) to find the average toxic loads for all regions. In this case, the summation is taken over the nodes belonging to the respective regions. We plot the toxic load corresponding to the tau protein for each region in Fig. \ref{fig:fig7}. The toxic load for the tau protein converges to different levels for different nodes due to the heterogeneous parameter values in the tau protein equation. According to the integrated data, the occipital region accumulates the most toxic concentration, followed by the parietal, basal ganglia, limbic, temporal, frontal, and brain stem. Moreover, the toxic propagation profile for each region is different. Some regions accumulate more toxic tau protein concentration after the initial progression of the disease but settle down to a comparatively lower concentration for a longer time, e.g., basal ganglia, parietal, and limbic. For the other regions, there is not much accumulation in the concentration after the start of the disease; rather, they slowly accumulate the toxic loads and help in disease progression. Figures \ref{fig:fig7}(\subref{fig:fig7a}) and (\subref{fig:fig7b}) show the toxic tau protein propagation in regions for the traditional non-fractional model and fractional model with $\alpha = 0.8$, respectively (other parameters are mentioned in the caption). This comparison demonstrates that the memory effect reduces the propagation speed in brain regions. We have observed the fractional model for other values of $\alpha (< 1)$, and the propagation speed decreases with decreasing values of $\alpha$.

\begin{figure}[ht!]
\centering
\includegraphics[width=\textwidth]{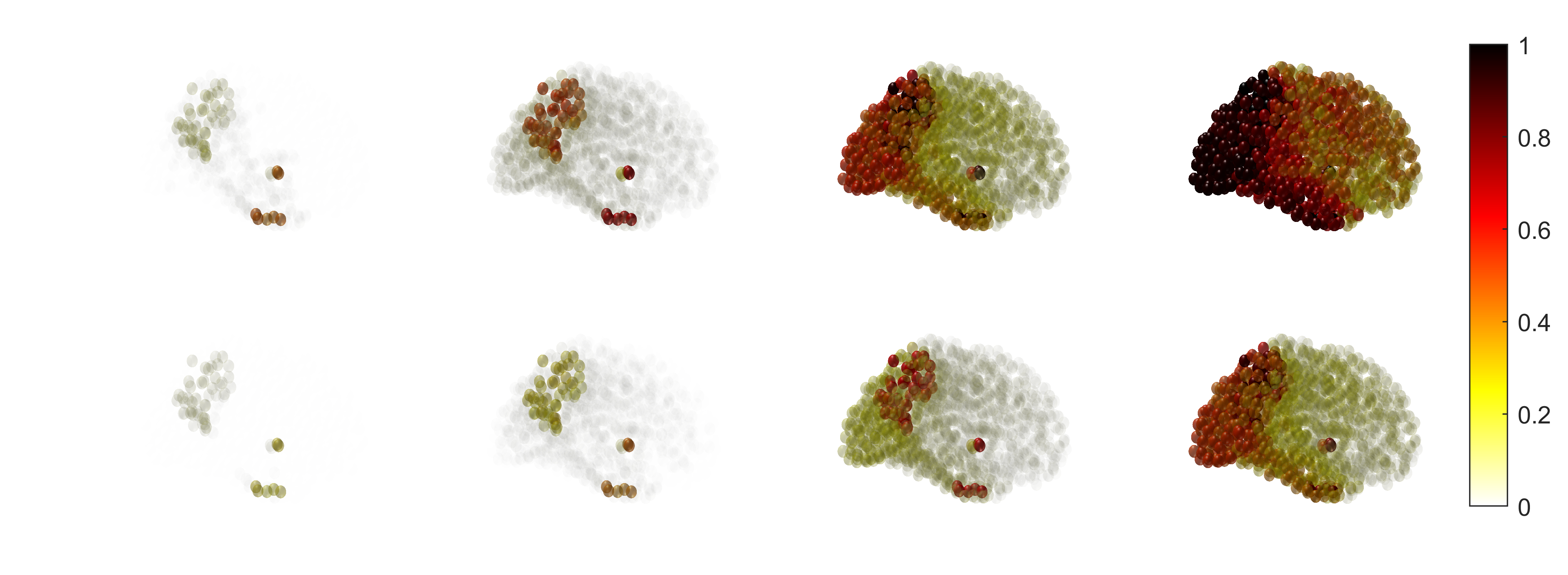}
\caption{(Color online) Node-wise damage propagation ($q$) for the traditional non-fractional (upper panel) and the fractional with $\alpha = 0.8$ (lower panel) models in brain connectome. The dark red represents the high damage, and the light yellow represents the low damage.}
\label{fig:fig8}
\end{figure}

We apply the same formula (\ref{ACF}) to find the regions' average damage profile. The damage profile for each region is different because of the non-uniform distributions of the toxic tau proteins. The region corresponding to the maximum toxic concentration is damaged first, then the region with the second-highest concentration, and so on. Figure \ref{fig:fig8} shows the node-wise neuronal damage propagation for non-fractional and fractional models. In the results, the damage dynamics are shown till $t = 400$ (non-dimensional time), but we have observed that the required time to damage each region in the brain for the fractional model is higher than the non-fractional model. Hence, the memory effect takes longer to damage the brain cells.

\section{Conclusions}{\label{sec6}}

This work uses a modified heterodimer model to explain interactions between two proteins A$\beta$ and $\tau$P. We have incorporated the coupled dynamics of astrocytes dynamics into the modified model and studied the dual role of astroglia before and after AD. Furthermore, we have studied the memory effect in AD progression, which is highly relevant to disease progression. Most of these investigations have been carried out by considering the heterogeneous parameter values, and it is a more realistic synthetic parameter set-up. 

Depending on the activated or deactivated astrocytes, the considered network model shows a dual behaviour in disease progression. The density of toxic amyloid-beta increases as the astrocyte clearance rate increases. For lesser astrocyte densities, however, the reverse scenario occurs. As a result, if enough astrocytes are present in the brain, they can reduce or postpone AD progression; otherwise, they help in AD progression. Furthermore, the fractional differential derivative framework presented here helps to model the memory effect on AD progression. We have shown that an increase in memory (by decreasing the parameter values $\alpha$) causes a delay in the toxic density propagations in the brain. As a result, it slows down AD progression in the brain, giving experimentalists more freedom in terms of parameters to fit their data appropriately.

We have studied the network model for the parameter values where primary and secondary tauopathy conditions are satisfied in distinct brain areas. This causes a non-homogeneous distribution of toxic tau proteins in the brain. In addition, the network model demonstrates that nodes with high connectivity have a higher chance of getting the infection and evolving into hubs for spreading disease. Furthermore, different neuronal damage profiles are shown on different brain IDs and in different brain regions. Hence, heterogeneous parameter values in the network model capture a realistic scenario of AD progression in the brain \cite{insel2020}. These non-uniform parameter values in the parameters involved in the amyloid-beta could be a good extension of this work. The coupling of astrocytes to A$\beta$ and $\tau$P represents an advancement in this direction, and one could use this model in an experimental configuration to improve data fitting. Along with the memory, considering heterogeneous parameter values corresponding to amyloid-beta or astrocytes in different brain IDs or regions is an important avenue for future research on this model. To a greater extent, neurodegenerative diseases involve complex and multiscale processes with multiple levels of biological framework, ranging from molecular and cellular to systemic and even societal. The presented work and the developed methodology allow us to reveal new trends and additional features of the underlying processes. More refined views on the complex dynamics of neurodegenerative diseases are expected with the subsequent incorporation of other scales (e.g., molecular-to-cellular, macro-to-micro) into the coupled biological framework.  

Validating the considered fractional-order derivative model requires robust datasets and experimental frameworks focusing on A$\beta$ and tau proteins. Some of the potential datasets and experimental frameworks can be used for future validation, and most of them have controlled access. NIAGADS is a collaboration between the National Institute on Ageing and the University of Pennsylvania that saves and distributes genetics and genomics data from AD, associated dementias, and ageing research to qualified researchers worldwide. The Alzheimer's Disease Neuroimaging Initiative (ADNI) provides complete data to research the course of A$\beta$ and tau pathology, including imaging, biomarker, and clinical evaluations. The AMP-AD Knowledge Portal provides free use of multi-omic data from Alzheimer's patients, including gene expression profiles for A$\beta$ and tau. Analyzing post-mortem brain tissue from Alzheimer's patients can also give the model real-world applicability.

A promising future direction involves the relationship between the memory effect at the molecular level and its phenotypic manifestations at the cellular and tissue levels \cite{okumura2023, mondal2024}. For example, biological mechanisms such as synaptic plasticity, protein misfolding, or feedback loops in tau and A$\beta$ propagation may explain the memory-like features in disease development. Data-driven simulation of network-based modelling approaches can offer deeper insights into the spatial and temporal patterns of tau deposition \cite{kim2024, vogel2023, ottoy2024}. In addition, data from neurobiological processes such as cellular signalling pathways, neuroinflammation, and neural connections, which are implicated in the spread of various disorders, may represent the model's history-dependent progression. The issue of higher-order interactions in complex networks such as human brain connectome is a very non-trivial task \cite{andjelkovic2020, tadic2024}. Nevertheless, focussing on identifying critical fractional orders that best fit experimental data and performing sensitivity analyses to highlight key regulatory factors can give more accurate models for predicting disease progression and developing targeted therapeutic strategies \cite{pini2025}.

\section{Methods}{\label{sec2}}


Alzheimer's disease strongly correlates with glial cells called astrocytes found in the central nervous system \cite{thuraisingham2022}. These astrocytes play a dual role in healthy and AD-affected brains, and in capturing such dynamics, we introduce an equation corresponding to astrocytes and modify the temporal model defined in \cite{thompson2020,pal2022} as
\begin{equation}\label{FMHDM}
\begin{aligned}
u_{t} & = u(a_{0} -a_{1}u) - a_{2}u\widetilde{u}, \\
\widetilde{u}_{t} & = -\widetilde{a}_{1}\widetilde{u} + a_{2}u\widetilde{u} -\mu \widetilde{u}(w-\widetilde{u}),\\
v_{t} & = v(b_{0} -b_{1}v) - b_{2}v\widetilde{v}-b_{3}\widetilde{u}v\widetilde{v},\\
\widetilde{v}_{t} & = -\widetilde{b}_{1}\widetilde{v} + b_{2}v\widetilde{v} +b_{3}\widetilde{u}v\widetilde{v},\\
w_{t} & = w(c_{0}-w/c_{1}),
\end{aligned}
\end{equation}
where the subscript $t$ denotes the first-order ordinary derivative with respect to $t$, with the initial conditions given by $u(0) = u_{0}$, $\widetilde{u}(0) = \widetilde{u}_{0}$, $v(0) = v_{0}$, $\widetilde{v}(0) = \widetilde{v}_{0}$, and $w(0) = w_{0}$. Here $u$ and $v$ are the healthy densities of A$\beta$ and $\tau$P, respectively, and $\widetilde{u}$ and $\widetilde{v}$ are the toxic densities A$\beta$ and $\tau$P, respectively. The parameters $a_{0}$ and $a_{1}$ denote the mean production and clearance rates of healthy A$\beta$, respectively, while $b_{0}$ and $b_{1}$ represent the mean production and clearance rates of healthy $\tau$P. The terms $\widetilde{a}_{1}$ and $\widetilde{b}_{1}$ describe the mean clearance rates of toxic forms of these proteins. The parameters $a_{2}$ and $b_{2}$ correspond to the mean conversion rates from healthy to toxic proteins. The coupling between the two proteins A$\beta$ and $\tau$P is captured by the parameter $b_{3}$. Finally, the variable $w$ represents the concentration of activated astrocytes, with $c_{0}$ as the production rate and $c_{0}c_{1}$ indicating the saturation point. The parameter $\mu$ is responsible for the dual role of astrocytes. If $w>\widetilde{u}$, then astrocytes clear the concentrations of the toxic amyloid-beta; otherwise, it helps to increase the toxic concentrations.


The reaction terms on the right side of the equation (\ref{FMHDM}) determine the substance concentration for any time $t>0$. Practically, it means that an individual who has had dementia for twenty years has the same chance of clearing Alzheimer’s as someone who had dementia less than ten years ago. It is an assumption based on the Markovian process, which is not generally valid. Non-Markovian processes have been playing an increasingly important role in studying living systems \cite{frank2013, aguilera2022, vilk2024}, and neuroscience research is no exception where such processes have to be incorporated in state-of-the-art models of neurodegenerative diseases. Clearly, the concentrations of the substances indicated above rely not only on the current time incident $t$ but also on the weighted average concentrations of the pastime range, say $[t_{p},t]$ for $t_{p}<t$. This is commonly referred to as the memory effect \cite{cressoni2012,cressoni2013,saeedian2017,mohammad2021,ghosh2021, wang2021, de2022}. The weight distribution relies on the power of the elapsed time, i.e., $(t-t_{p})$, and follows the power law correlation function \cite{stanislavsky2000,saeedian2017}. We can select $t_{p} = 0$ without loss of generality. Now, incorporating these into the mathematical model (\ref{FMHDM}), we obtain the fractional order differential equations as
\begin{equation}\label{MHDMPL}
\begin{aligned}
D^{\alpha}_{t}u& = u(a_{0} -a_{1}u) - a_{2}u\widetilde{u}, \\
D^{\alpha}_{t}\widetilde{u} & = -\widetilde{a}_{1}\widetilde{u} + a_{2}u\widetilde{u} -\mu \widetilde{u}(w-\widetilde{u}),\\
D^{\alpha}_{t}v & = v(b_{0} -b_{1}v) - b_{2}v\widetilde{v}-b_{3}\widetilde{u}v\widetilde{v},\\
D^{\alpha}_{t}\widetilde{v} & = -\widetilde{b}_{1}\widetilde{v} + b_{2}v\widetilde{v} +b_{3}\widetilde{u}v\widetilde{v},\\
D^{\alpha}_{t}w & = w(c_{0}-w/c_{1}),
\end{aligned}
\end{equation}
where $D^{\alpha}_{t}z(t)$ stands for the Caputo fractional derivative, defined as
$$D^{\alpha}_{t}z(t) = \frac{1}{\Gamma (1-\alpha)} \int_{0}^{t} \frac{z'(s)}{(t-s)^{\alpha}}ds,~~0<\alpha <1,$$
and $z'$ denotes the first-order ordinary derivative of $z$. Here in the modified fractional differential model (\ref{MHDMPL}), the influence of memory decreases when $\alpha \rightarrow 1$, and the system tends toward a memoryless system \cite{saeedian2017, caputo1967}. In addition, the accumulations of toxic amyloid-beta and tau proteins cause neuronal damage. We consider the memory effect in such neuronal damage equation by modelling it by the following equation:
\begin{equation}{\label{TDE}}
    D^{\alpha}_{t}q = (1-q)(k_{1}\widetilde{u} + k_{2}\widetilde{v} + k_{3}\widetilde{u}\widetilde{v} +k_{4}q),
\end{equation}
with a non-negative initial condition $q(0) = q_{0}$. The case $q=0$ signifies a healthy state, i.e., neurons are properly functioning, and $q=1$ implies an unhealthy or no longer functioning state \cite{thompson2020}. In studying neurodegenerative diseases, the development of coupled dynamic models plays a critical role. Different aspects of such coupled models, including those at the neuron-glial level and the toxic amyloid-beta dynamics accounting for astrocytes, have been studied in recent papers \cite{pal2022a, pal2022b, shaheen2023, pal2023, shaheen2023a, shaheen2023b}. The present work is a new step in further refining such coupled models where the Markovian assumption, which cannot be justified in the general dynamic studies of neurodegenerative diseases, is removed. Before proceeding to the analysis of such refined models, we note that such models degenerate into the Markovian case once, in the neural damage equation presented above, the fractional derivative $D^{\alpha}_{t}q$ is replaced by the ordinary derivative $dq/dt$  \cite{thompson2020,pal2022}.

\subsection{Analysis for the Homogeneous System}{\label{sec3}}

Here, we analyze the time-varying behaviour of both the fractional and non-fractional models. First, we describe the equilibria of the non-fractional model (\ref{FMHDM}) and their stability behaviours. These equilibria for the system (\ref{FMHDM}) are the time-independent solutions of (\ref{FMHDM}), and they can be obtained by solving the system (\ref{FMHDM}) with the vanishing time derivatives. In addition, they depend on the parameter values, and we calculate them numerically later on. Moreover, each equilibrium point's stability is determined by the nature of all the eigenvalues of the Jacobian matrix calculated at that point. For any equilibrium point $E_{*} = (u_{*},\widetilde{u}_{*},v_{*},\widetilde{v}_{*},w_{*})$, the Jacobian matrix of the system (\ref{FMHDM}) is given by
$$\mathbf{J}_{*} = 
\begin{pmatrix}
    a_{11} & a_{12} & 0 & 0 & 0\\
    a_{21} & a_{22} & 0 & 0 & a_{25}\\
    0 & a_{32} & a_{33} & a_{34} & 0\\
    0 & a_{42} & a_{43} & a_{44} & 0\\
    0 & 0 & 0 & 0 & a_{55}\\
\end{pmatrix},$$
where $a_{11} = a_{0}-2a_{1}u_{*}-a_{2}\widetilde{u}_{*}$, $a_{12} = -a_{2}u_{*}$, $a_{21} = a_{2}\widetilde{u}_{*}$, $a_{22} = -\widetilde{a}_{1}+a_{2}u_{*}-\mu (w_{*}-2\widetilde{u}_{*})$, $a_{25} = -\mu \widetilde{u}_{*}$, $a_{32} = -b_{3}v_{*}\widetilde{v}_{*}$, $a_{33} = b_{0}-2b_{1}v_{*}-b_{2}\widetilde{v}_{*}-b_{3}\widetilde{u}_{*}\widetilde{v}_{*}$, $a_{34} = -b_{2}v_{*}-b_{3}\widetilde{u}_{*}v_{*}$, $a_{42} = b_{3}v_{*}\widetilde{v}_{*}$, $a_{43} = b_{2}\widetilde{v}_{*}+ b_{3}\widetilde{u}_{*}\widetilde{v}_{*}$, $a_{44} = -\widetilde{b}_{1}+b_{2}v_{*}+b_{3}\widetilde{u}_{*}v_{*}$, and $a_{55} = c_{0}-2w_{*}/c_{1}$. If the real components of all the eigenvalues of $\mathbf{J}_{*}$ are negative, then the equilibrium point $E_{*}$ is stable; otherwise, it is unstable. In addition, the non-fractional damage equation has only one equilibrium point $q_{*} = 1$, which is stable. Furthermore, all the equilibrium points for the traditional non-fractional model are also the equilibrium points for the fractional model, but their stability behaviours are not the same for both models. For the fractional model with fixed $\alpha$, an equilibrium point $E_{*}$ is stable if all the eigenvalues $\lambda_{i}$ ($i=1,\ldots, 5$) of $\mathbf{J}_{*}$ satisfy $|\arg (\lambda_{i})|>\alpha\pi/2$; otherwise, it is unstable \cite{ghosh2021}.

\subsection{Network Model in the Brain Network}{\label{sec4}}

Before going to the brain connectome network model, we extend the temporal model (\ref{MHDMPL}) into the reaction-diffusion model in a subset of the Euclidean space. This spatial extension is crucial in understanding the spatio-temporal evolution of A$\beta$ and $\tau$P in the brain connectome. Indeed, several vivo and vitro studies indicated that the tau protein aggregates and can propagate along synapsis \cite{devos2018}. A spatio-temporal extension of the fractional model (\ref{MHDMPL}) in a general continuous spatial domain $\Omega\subset\mathbb{R}^{3}$ is given by
\begin{equation}\label{SMHDMPL}
\begin{aligned}
D^{\alpha}_{t}u& = \nabla\cdot (\mathbf{D}_{1} \nabla u) + u(a_{0} -a_{1}u) - a_{2}u\widetilde{u}, \\
D^{\alpha}_{t}\widetilde{u} & = \nabla\cdot (\widetilde{\mathbf{D}}_{1} \nabla\widetilde{u}) -\widetilde{a}_{1}\widetilde{u} + a_{2}u\widetilde{u} -\mu \widetilde{u}(w-\widetilde{u}),\\
D^{\alpha}_{t}v & = \nabla\cdot (\mathbf{D}_{2} \nabla v) + v(b_{0} -b_{1}v) - b_{2}v\widetilde{v}-b_{3}\widetilde{u}v\widetilde{v},\\
D^{\alpha}_{t}\widetilde{v} & = \nabla\cdot (\widetilde{\mathbf{D}}_{2} \nabla\widetilde{v}) -\widetilde{b}_{1}\widetilde{v} + b_{2}v\widetilde{v} +b_{3}\widetilde{u}v\widetilde{v},\\
D^{\alpha}_{t}w & = w(c_{0}-w/c_{1}).
\end{aligned}
\end{equation}
The first term on the right-hand side of the first four equations accounts for the random movement of concentrations in the domain $\Omega$. It is assumed that the density of astrocytes is homogeneous in the domain $\Omega$. Here, $\mathbf{D}_{1}$, $\widetilde{\mathbf{D}}_{1}$, $\mathbf{D}_{2}$, and $\widetilde{\mathbf{D}}_{2}$ are the diffusion tensors which describe each protein's spreading. We consider the same damage equation (\ref{TDE}) in this spatial extension, and hereafter, the damage $q$ also depends on the spatial location, i.e., $q(\mathbf{x},t), \mathbf{x}\in \Omega$. The astrocytes also affect the dynamics of the neurons, as they are implicitly involved through toxic amyloid beta.

The main goal is to study the disease progression within the brain connectome. The modified model (\ref{SMHDMPL}) is defined in a continuous domain $\Omega$. Now, we develop a network mathematical model that correlates with the model (\ref{SMHDMPL}) so that we can integrate the brain connectome data \cite{thompson2020,pal2022}. Suppose $\mathbf{G}$ is the network brain data with $V$ nodes and $E$ edges. We generate the adjacency matrix $\mathbf{A}$ for the graph $\mathbf{G}$, which enables us to build the graph's Laplacian. The $(i,j)$ ($i,j=1,2,3,\ldots,N$) element of the matrix $\mathbf{A}$ is defined as follows: $$A_{ij}=\frac{n_{ij}}{l_{ij}^2},$$ where $l_{ij}^{2}$ represents the mean length squared between the nodes $i$ and $j$ and $n_{ij}$ is the mean fiber number. Let us now define the elements of the Laplacian matrix $\mathbf{L}$ as $$L_{ij}=\rho (D_{ij}-A_{ij}),~~i,j=1,2,3,\ldots,N,$$ where $D_{ii}=\sum_{j=1}^{N}A_{ij}$ are the elements of the diagonal weighted-degree matrix and $\rho$ is the diffusion coefficient. This Laplacian matrix is used to construct a network model for graph $\mathbf{G}$. We employ superscript notations for their respective Laplacian matrices to differentiate the diffusibility of each protein in the brain. In this case, the dynamics of each node $j (j=1,2,3,\ldots,N)$ can be obtained by:
\begin{equation}\label{NSMHDMPL}
\begin{aligned}
D^{\alpha}_{t}u_{j}& = -\sum_{k=1}^{N}L_{jk}^{u}u_{j} + u_{j}(a_{0} -a_{1}u_{j}) - a_{2}u_{j}\widetilde{u}_{j}, \\
D^{\alpha}_{t}\widetilde{u}_{j} & = -\sum_{k=1}^{N}L_{jk}^{\widetilde{u}}\widetilde{u}_{j} -\widetilde{a}_{1}\widetilde{u}_{j} + a_{2}u_{j}\widetilde{u}_{j} -\mu \widetilde{u}_{j}(w_{j}-\widetilde{u}_{j}),\\
D^{\alpha}_{t}v_{j} & = -\sum_{k=1}^{N}L_{jk}^{v}v_{j} + v_{j}(b_{0} -b_{1}v_{j}) - b_{2}v_{j}\widetilde{v}_{j} -b_{3}\widetilde{u}_{j}v_{j}\widetilde{v}_{j},\\
D^{\alpha}_{t}\widetilde{v}_{j} & = -\sum_{k=1}^{N}L_{jk}^{\widetilde{v}}\widetilde{v}_{j} -\widetilde{b}_{1}\widetilde{v}_{j} + b_{2}v_{j}\widetilde{v}_{j} +b_{3}\widetilde{u}_{j}v_{j}\widetilde{v}_{j},\\
D^{\alpha}_{t}w_{j} & = w_{j}(c_{0}-w_{j}/c_{1}),
\end{aligned}
\end{equation}
the corresponding damage equation can be obtained by the fractional differential equation:
\begin{equation}{\label{NTDE}}
    D^{\alpha}_{t}q_{j}  = (1-q_{j})(k_{1}\widetilde{u}_{j} + k_{2}\widetilde{v}_{j} + k_{3}\widetilde{u}_{j}\widetilde{v}_{j} +k_{4}q_{j}),
\end{equation}
with non-negative initial conditions. The equilibria of the homogeneous system correspond to the network model's homogeneous stationary steady-states (\ref{NSMHDMPL}). 

\section*{Acknowledgements}
The authors are grateful to the NSERC and the CRC Program for their support. RM also acknowledges the support of the BERC 2022-2025 program and the Spanish Ministry of Science, Innovation and Universities through the Agencia Estatal de Investigacion (AEI) BCAM Severo Ochoa excellence accreditation SEV-2017-0718 and the Basque Government fund AI in BCAM EXP. 2019/00432. This research was enabled in part by support provided by SHARCNET \url{(www.sharcnet.ca)} and Digital Research Alliance of Canada \url{(www.alliancecan.ca)}.

\bibliographystyle{elsarticle-num}
\bibliography{References}

\end{document}